\documentclass[prl,twocolumn,amsmath,amssymb]{revtex4-2}

\usepackage{graphicx}
\usepackage{bm}
\usepackage{physics}
\usepackage{color}
\newcommand{\s}{\xi}

\newcommand{\oJ}{\hat{\mathcal{J}}}

\begin{document}
\title{Manipulating intertwined orders in solids with quantum light}
\author{Jiajun Li}
\affiliation{Institute of Theoretical Physics, University of Erlangen-Nuremberg, 91052 Erlangen, Germany}
\author{Martin Eckstein}
\affiliation{Institute of Theoretical Physics, University of Erlangen-Nuremberg, 91052 Erlangen, Germany}

\date{\today}

\begin{abstract}
Intertwined orders exist ubiquitously in strongly correlated electronic systems and lead to intriguing phenomena in quantum materials. In this paper, we explore the unique opportunity of manipulating intertwined orders through entangling electronic states with quantum light. Using a quantum Floquet formalism to study the cavity-mediated interaction, we show the vacuum fluctuations effectively enhance the charge-density-wave correlation, giving rise to a phase with entangled electronic order and photon coherence, with putative superradiant behaviors in the thermodynamic limit. Furthermore, upon injecting even one single photon in the cavity, different orders, including $s$--wave and $\eta$--paired superconductivity, can be selectively enhanced. Our study suggests a new and generalisable pathway to control intertwined orders and create light-matter entanglement in quantum materials. The mechanism and methodology can be readily generalised to more complicated scenarios.
\end{abstract}

\maketitle

\textbf{Introduction.} Two condensed matter phases are referred as ``competing" if favoring one automatically suppresses the other. In quantum materials, this situation is often related to the interwining of multiple orders, when the competing orders correspond to orthogonal directions in a larger order-parameter space, and are sensitive to small parameter changes. A paradigmatic example of competing phases is given by intertwining of charge-density waves (CDW), characterized by a staggered pattern of charge occupation, and superconducting (SC) phases  \cite{kivelson2003,ghiringhelli2012,chang2012,tranquada1995,sipos2008,fradkin2015}, which is minimally described by the attractive Hubbard model \cite{micnas1990}. 

A promising pathway of controlling these orders is to dress materials with strong laser fields, termed Floquet engineering \cite{basov2017,bukov2015,eckardt2017,oka2019}. The stability of CDW and SC phases is shown to be selectively controlled by classical-light driving \cite{kitamura2016,sentef2017}. Other intriguing scenarios include light-induced superconductivity and anomalous quantum Hall effect \cite{wang2013,mciver2020,buzzi2020}. In these cases, the \emph{quantum fluctuation} of electromagnetic fields is usually negligible. However, the ultrastrong light-matter coupling (USC) is recently realized in cavity quantum electrodynamics (QED) \cite{kockum2019}, where the quantum fluctuations become dominant and can entangle with different macroscopic states  \cite{julsgaard2001}. This opens up the unique possibility to control intertwined orders in the hybrid light-matter phases \cite{smolka2014,thomas2019,sentef2018,mazza2019,schlawin2019,schlawin2019b,wang2019,li2020,lenk2020,gao2020,ashida2020,rohn2020}. 

In spite of a clear analogy between classical electrodynamics and cavity QED \cite{schaefer2018,sentef2020}, a few-photon state in a weakly driven cavity, with strong quantum fluctuations, differs dramatically from classical light in free-space, and does not necessarily control the material properties in a similar manner to the Floquet engineering \cite{kiffner2019, mentink2015}. A systematic theory of the cavity-coupled solids for such excited or driven cavity states is therefore interesting, but still at its infancy. In this paper, we use a quantum Floquet formalism to examine the possibility of controlling competing phases by creating highly entangled electronic and photon states in quantum materials. We show that the ground state of the cavity-coupled attractive Hubbard model features entangled electronic order and photon coherence, with an enhancement of the charge density wave (CDW) order. With appropriate protocols, it is possible to selectively enhance CDW, $s$--wave and even $\eta$--pairing superconductivity (SC) \cite{yang1990} by creating more photons in the cavity. The conclusions are confirmed with the exact diagonalization of 1D Hubbard chains. 

\textbf{Quantum Floquet formalism.} 
We consider a half-filled attractive Hubbard model placed in a cavity. The cavity contains a single photon mode with polarization $\bm e_p$. For simplicity we consider the 1D case, with $\bm e_p$ parallel to the chain. (The formalism presented below is independent of dimensionality.) The Hamiltonian reads
\begin{align}
    \hat{H}=-t_0\sum_{\langle ij\rangle\sigma} e^{i\hat{\phi}_{ij}}c^\dag_{i\sigma} c_{j\sigma} - U \sum_{i} 
n_{i\uparrow}n_{i\downarrow}
+\Omega 
a^\dag a,
\label{Ho}
\end{align}
where $\hat{\phi}_{ij}=\bm A\cdot \bm d_{ij}=g\xi_{ij}(a+a^\dag)$ is the Peierls phase with bond dipole $\bm d_{ij}$ and vector potential $\bm A = \bm e_p A_0(a+a^\dag)$, i.e., $\s_{ij}=1$ for hopping parallel to the polarization and is $-1$ for the anti-parallel direction,
and $g=|\bm d_{ij}|A_0$ is the dimensionless coupling parameter. The corresponding electric field is, as usual, $\bm E=i\bm e_p \Omega A_0(a-a^\dag)$. Note that the above Hamiltonian is explicitly 
gauge-invariant 
and retains all the higher-order coupling terms, including the so-called diamagnetic term ($\bm A^2$)
\cite{li2020}. A simple truncation of the coupling can lead to unphysical results \cite{kiffner2019b,andolina2019}.

For $g = 0$, the model allows for three intertwined orders: the commensurate CDW featuring staggered electron occupations, the uniform or $s$--wave SC, and $\eta$--paired SC \cite{kitamura2016}. The latter has a staggered pair-field amplitude, but nevertheless shows superconducting properties such as the Meissner effect \cite{li2019}.  \emph{Degenerate} CDW and $s$--wave SC orders compete in the ground state due to the $SO(4)$ symmetry. The fate of the intertwined CDW and SC orders is altered by the \emph{cavity-mediated interaction}, which is studied in a Floquet-like formalism below. The latter is similar in spirit as some recent works \cite{sentef2020, schaefer2018,kiffner2019}, but will be formulated more explicitly in a photon-number basis. We expand the Hamiltonian~\eqref{Hfl} in the photon number basis $\hat{H} = \sum_{nm} (\mathbb{I}_{\rm el}\otimes \ket{n}\bra{n}) \hat{H} (\mathbb{I}_{\rm el}\otimes\ket{m}\bra{m}) = \mathcal{H}_{nm}\otimes\ket{n}\bra{m}$, where $\mathbb{I}_{\rm el}$ is the identity operator in the electronic Hilbert space, and introduce the \emph{quantum} Floquet matrix 
\begin{align}
   &\mathcal{H}_{nm}=\mathcal{H}^0_{nm}+(-U \sum_{i} n_{i\uparrow}n_{i\downarrow} + n\Omega)\delta_{nm}\nonumber\\
  &\text{with } \mathcal{H}^0_{nm}=t_0\sum_{\langle ij\rangle\sigma} i^{|n-m|}\s_{ij}^{n-m}j_{n,m}c^\dag_{i\sigma} c_{j\sigma},
    \label{Hfl}
\end{align} 
where $\langle n| e^{i\hat{\phi}_{ij}}|m\rangle=i^{|n-m|}\s_{ij}^{n-m}j_{n,m}$ represent the matrix elements of the Peierls 
phase. They can be evaluated as a \emph{finite} sum \footnote{see Supplemental Material},
\begin{align}
	j_{n,m} = e^{-g^2/2}\sum_{k=0}^m\frac{(-1)^kg^{2k+|n-m|}}{k!(k+|n-m|)!}\sqrt{\frac{n!}{m!}}\frac{m!}{(m-k)!},
\end{align}
for $n>m$ and $j_{n,m}=j_{m,n}$. The expression for $\mathcal{H}_{nm}$ resembles the Floquet matrix Hamiltonian \cite{tsuji2008}, but unlike the latter it is not translationally invariant in the photon index ($\mathcal{H}^0_{nm}\ne\mathcal{H}^0_{n+\ell, m+\ell}$), and the indices are restricted to $n,m\geq 0$. Nevertheless, in the semi-classical limit $n,m\to\infty$ with $g\sqrt{n}$ finite \cite{sentef2020}, $j_{n,m}$ converges to the Bessel function $J_{|n-m|}(2g\sqrt{n})$, so that Eq.~\eqref{Hfl} recovers the Floquet Hamiltonian \cite{Note1}. Moreover, similar to the Bessel functions, the function $j_{n,m}$ decay super-exponentially as $|n-m|\to\infty$, which decouples the quantum Floquet bands for large $|n-m|$ and allows for efficient numerical evaluations. 

\begin{figure}
\includegraphics[scale=0.6]{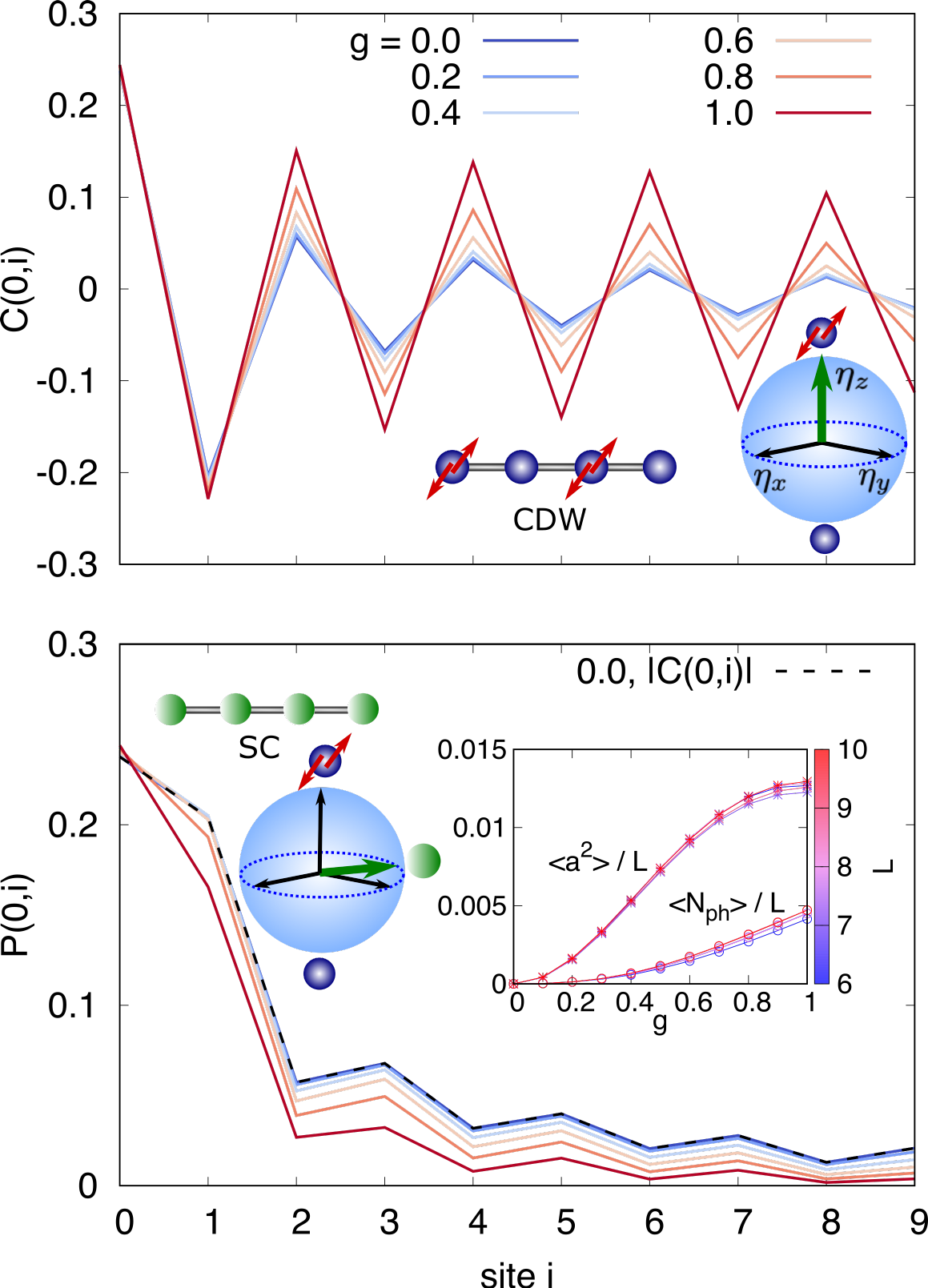}
\caption{Spatial correlation of charge ($C(0,i)$) and superconducting ($P(0,i)$) order for increasing photon coupling $g$, for  $U = 8.0t_0$ and $\Omega = 6.0t_0$. The dashed line in the lower panel shows the absolute value $|C(0,i)|$ at $g=0$ for comparison. The inset shows the photon occupation $N_{ph}=\langle a^\dag a\rangle$ and $\langle a^2\rangle$, scaled with $L=6,\ldots,10$. The figures also schematically show the CDW and SC orders represented by the green arrow on the Bloch sphere in the $\eta$--pseudospin space.}
\label{gs}
\end{figure}

\textbf{Strong coupling expansion.} 
The quantum Floquet Hamiltonian, similar to the classical Floquet approach, provides an intuitive picture of the underlying physics. For example, the hopping of an electron can result in a shift of the quantum Floquet index $n$, corresponding to the emission or absorption of photons. Technically, this allows for a systematic strong coupling expansion. Assuming off-resonance $\Omega\ne U$, the low-energy physics of \eqref{Hfl} in the limit $U\gg t$ can be captured by an effective pseudospin model \cite{micnas1990, kitamura2016} from a Schrieffer-Wolff transformation. When projected to a given photon number sector $n$, the effective Hamiltonian reads $H^{\rm eff}_{nn} = \sum_l \mathcal{P}_0\mathcal{H}^0_{n,n+l}\mathcal{P}_1\mathcal{H}^0_{n+l,n}\mathcal{P}_0/(U+l\Omega)$, where $\mathcal{P}_i$ is the projection operator to the subspace of $i$ electronic excitations. One obtains 
\begin{align}
    H^{\rm eff}_{nn}=\frac{1}{2}J^{SC}_{\rm ex}\sum_{\langle ij\rangle}(\eta^+_i\eta^-_j + {\rm h.c.}) + J^{CDW}_{\rm ex}\sum_{\langle 
ij\rangle}
    \eta^z_i\eta^z_j,
    \label{Heff}
\end{align}
where the $\eta$--pseudospin is defined as 
\begin{align}
    \eta^+_i&=(\eta^-_i)^\dag=(-1)^ic^\dag_ic^\dag_j,\nonumber\\
    \eta^z_i&=(n_i - 1) / 2,
\end{align}
so pseudospin $\eta^{\pm}$ represents pairing and $\eta^z$ corresponds to charge, see the sketch in Fig.~\ref{gs}. In the uncoupled ($g=0$) case $J^{SC}_{\rm ex}=J^{CDW}_{\rm ex}=J_{\rm ex} = 2t_0^2/U$. For $g\neq 0$, the exchange coupling contains contributions from all virtual hopping processes with intermediate states in different photon-number sectors (labeled by $l$) and the processes associated with $J^{SC}$ and $J^{CDW}$ capture different phase factors \cite{kitamura2016}, 
\begin{align}
 \left\{ 
\begin{array}{c}
J^{SC}_{\rm ex} 
\\
 J^{CDW}_{\rm ex}
 \end{array}
\right\}
&= J_{\rm ex} \sum_{l\ge-n}^\infty
\left\{ 
\begin{array}{c}
(-1)^l 
\\
+1
\end{array}
\right\}
\frac{j_{n,n+l}j_{n+l,n}}{1+l\Omega/U}.
    \label{jj}
\end{align}
The full strong-coupling model also contains a pseudospin-photon coupling which is off-diagonal in the photon number (see below), but for $\Omega\gg J_{\rm ex}$, transitions between photon sectors are suppressed and the electronic configuration is determined by Eq.~\eqref{Heff} for fixed $n$.

\textbf{Entangling orders with vacuum fluctuations.} 
In the cavity ground state ($n=0$) the induced interaction exclusively enhances $J^{CDW}$ and suppresses $J^{SC}$ irrespective of the values of $U$ and $\Omega$, because $j_{l,0}=e^{-g^2/2}g^l/\sqrt{l!}>0$. The relevant factor $e^{-g^2/2}$ is due to the cavity-induced dynamical localization \cite{sentef2020}. This behavior is dramatically different from classical Floquet driving, where a blue-detuned light ($\Omega > U$) enhances superconductivity \cite{kitamura2016, sentef2017, fujiuchi2020}. To confirm this prediction from the
effective pseudospin model, we solve the original Hamiltonian \eqref{Ho} using exact diagonalization (ED). The ground state is obtained with the Lanczos algorithm, assuming half-filling and $\hat{S}_z=0$. The trend of forming CDW and SC orders is reflected by the charge and pairing correlation functions $C(0,i)=\frac{1}{4}\langle (n_0-1) (n_i-1) \rangle$ and $P(0,i)=\frac{1}{2}\langle c^\dag_{0\uparrow}c^\dag_{0\downarrow}c_{i\downarrow}c_{i\uparrow}\rangle$, see Fig.~\ref{gs} for $L=10$ under open boundary condition. At $g=0$, both functions have identical magnitude. As $g$ increases, a staggered charge correlation is continuously enhanced, corresponding to the enhanced CDW order, while the decreased pairing correlation indicates suppressed SC order. The same qualitative behavior is observed for $\Omega > U$, although the effect is weaker due to a larger denominator $1/(U+l\Omega)$ in Eq.~\eqref{jj}. This confirms our analytic theory.  

\begin{figure}
\includegraphics[scale=0.37]{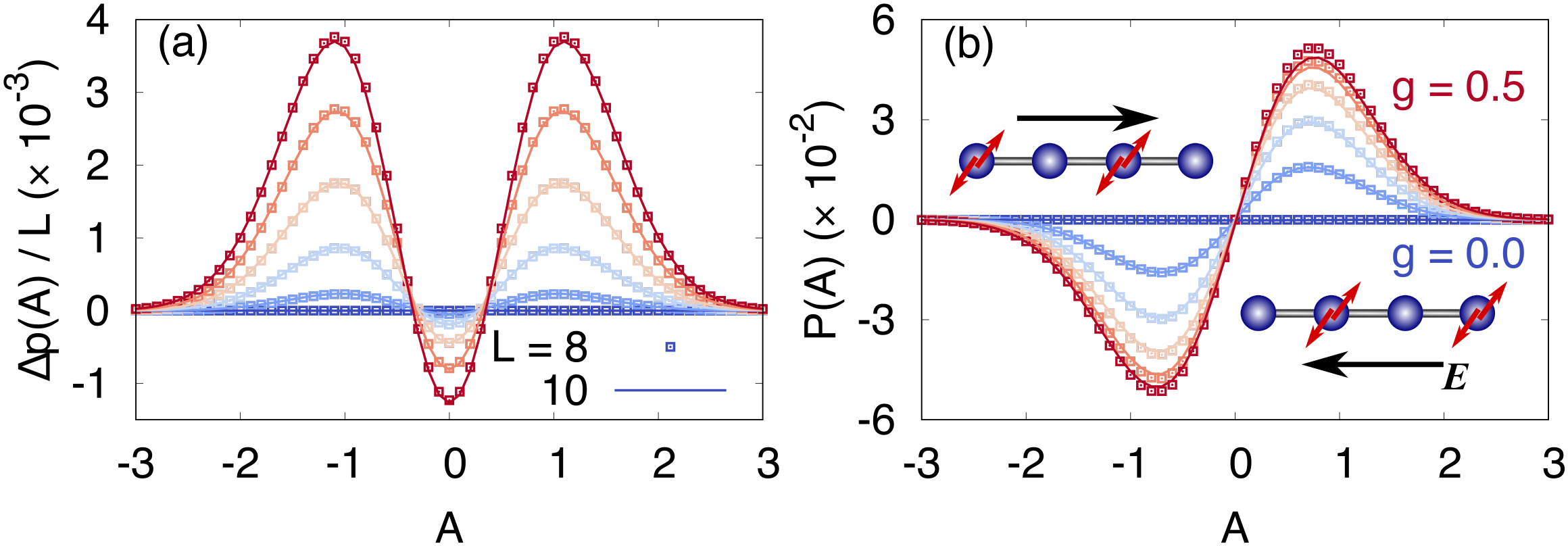}
\caption{Entanglement of the electronic order and photon coherent states. (a) Difference of probability $p(A)=\operatorname{Tr}(\rho_G \Pi_{A})$ for varying $g$ and the uncoupled case ($g=0$). $\rho_G=\ket{GS}\bra{GS}$ is the ground state density matrix. The values are normalized by $1/L$. (b) Charge polarization when measured in the projected state, i.e. $P(A)=\operatorname{Tr}[(n_{0}-n_{L-1})\rho_G \Pi_{A}]$.  Colors from blue to red indicate coupling $g=0.0,0.1,\ldots,0.5$. The sketch shows the CDW configurations corresponding to the peaks.}
\label{etg}
\end{figure}

Another intriguing aspect is the emergent light-matter mixing. Indeed, the photon occupation $N_{ph}=\langle a^\dag a\rangle$ scales almost linearly with system size $L$ (Fig.~\ref{gs} inset), implying a macroscopic $\langle a^\dag a\rangle\sim L$ in the thermodynamic limit, or a superradiant phase \cite{viehmann2011}. In the strong coupling picture, the light-matter entangling comes from two facts: (i) In the Schrieffer-Wolff transformation, photon operators are dressed, and the photon number $n$ in Eq.~\eqref{Heff} differs from the bare $\langle a^\dagger a\rangle$. The non-zero $\langle a^2\rangle$ shows that the dressed zero-photon state has some squeezed character (though $\langle a\rangle=0$). 
(ii) Moreover, when we restore the photon operators in the Hamiltonian \cite{sentef2020}, up to first order in $g$ one gets a (somewhat expected) Dicke-type coupling {$g_{\rm eff}i(a-a^\dag)\sum_{\langle ij\rangle}\xi_{ij}(n_i-n_j)$}, where $g_{\rm eff}$ is of order $gt_0^2/U$.  For an open chain the total charge polarization $P=n_0-n_{L-1}$ therefore couples to the electric field $ig(a-a^\dag)$ \cite{sentef2017,felicetti2020}. CDW configurations with $P>0$ ($n_0=2,n_{L-1}=0$ in the extreme case) and $P<0$ thus entangle with the photon states of $\langle \bm E\rangle\propto\pm \bm e_p$, which explains the behavior observed in Fig.~\ref{etg}.

The light-matter entangling can be highlighted in an intriguing manner by analyzing a projective measurement of the electric field amplitude (implemented by a projection $\Pi_A=\ket{iA}\bra{iA}$ on a coherent state of amplitude $iA$): At $g>0$, the probability distribution for the field acquires a double peak structure with maxima at $A\sim\pm1$ (Fig.~\ref{etg}a), and the matter is left in states of different charge polarization $P$  depending on the outcome of the measurement (Fig.~\ref{etg}b). This is impossible for a product state where measuring the photon would leave electrons unaffected.  While the global system does not break the symmetry ($\langle a\rangle=0$), the light-matter wave function has its weight centered at two semi-classical configurations with $A\approx1,P>0$ and $A\approx-1,P<0$ (see sketch in panel (b)). The superposition may collapse to a superradiant order under decoherence \footnote{Roughly speaking, the entangled state can be represented by $\ket{P>0}\ket{A\approx1}\pm \ket{P<0}\ket{A\approx-1}$. Note that in the language of macroscopic electrodynamics, the field $i(a-a^\dag)$ of amplitude $iA$ actually corresponds to the displacement field}.

\begin{figure}
\includegraphics[scale=0.8]{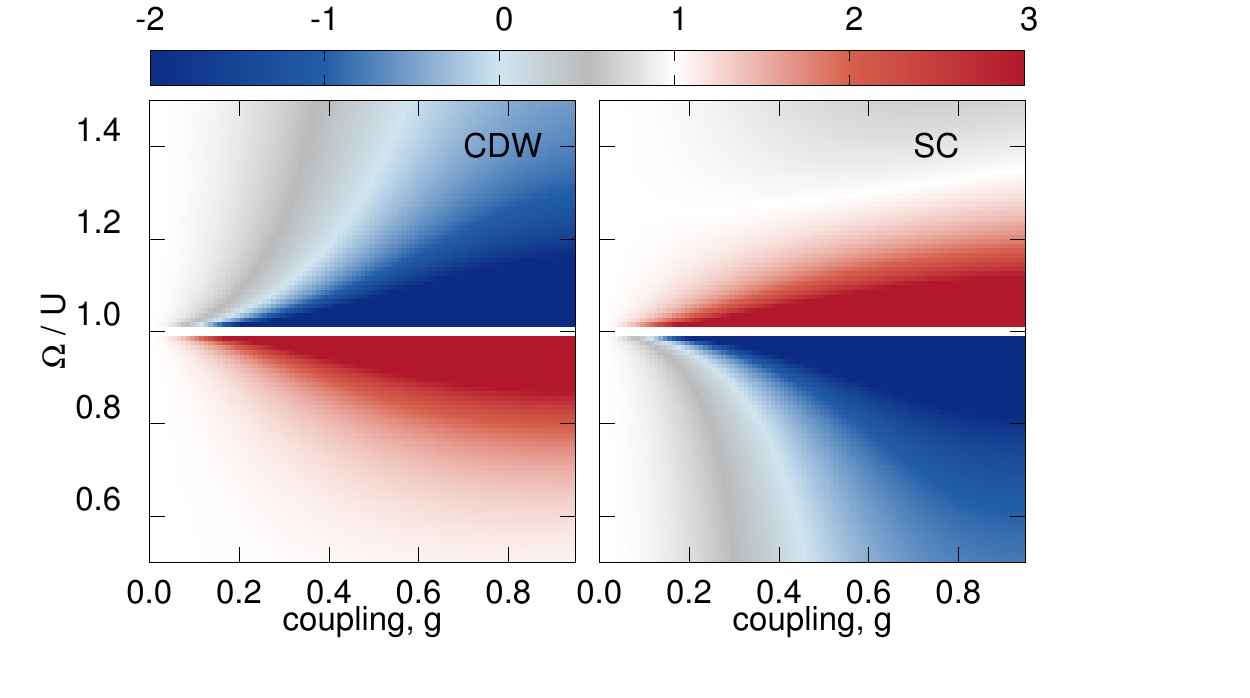}
\caption{The selective enhancement of different orders in the presence of one photon. The color represents the value of $J^{SC}_{\rm ex}$ and $J^{CDW}_{\rm ex}$ normalized by the uncoupled $J_{\rm ex}(0) = 2t_0^2/U$. The exchange coupling is enhanced in the red region and suppressed in the grey region. In particular, the exchange coupling changes its sign in the blue region.}
\label{oneph}
\end{figure}

\textbf{Enhanced SC in the few-photon regime.} To explore the possibility of selectively enhancing different orders, we now turn to the case of a driven cavity. Physically, we address this regime by injecting a finite number $n$ of photons into the cavity. The key difference between $n=0$ and $n>0$ in the couplings Eq.~\eqref{jj} is the existence of intermediate states with $l< 0$ (photon absorption), which contribute \emph{negative} denominators $1+l \Omega/U$. Even the presence of a single photon allows the selective enhancement of CDW and SC orders, see Fig.~\ref{oneph}. In general, the CDW is enhanced in the red-detuned regime, while the SC is enhanced in the blue-detuned regime. More interestingly, there is a wide regime (though being close to the resonance $\Omega\sim U$) where the exchange coupling changes its sign. In this case, a negative $J^{SC}_{\rm ex}$ favors the staggered, or $\eta$--paired superconductivity \cite{rosch2008,kitamura2016,li2019}, and a negative $J^{CDW}_{\rm ex}$ leads to a trend of \emph{charge segregation}, where doublons tend to stick together and repel holons. The same qualitative physics is found for more photons $N_{ph} \ge 2$. Note a Fock state with fixed $n$ has zero coherent amplitude, with no classical counterpart. In particular, superconductivity is enhanced by weak quantum light close to a Fock state ($n\gtrsim 1)$, but \emph{not} by classical light (coherent state) of similar amplitude, close to vacuum. 

In a real experiment, the multi-photon regime is realized through driving with an external laser field, which does not necessarily lead to a Fock state. However, the fine control of cavity photon number is supported by the strong non-linear effects of light-matter coupling. Specifically, the injection of one photon into the cavity modifies the ``internal state" of the cavity-matter system, changing the energy cost of injecting a second photon (appendix). The external driving can, therefore, be made resonant with selected photon numbers. The preparation of definite Fock states may be delicate, but the enhancement of  SC order remains robust even for a superposition of multi-photon states, because for given $\Omega/U$ it shows the same trend for different $n>0$ (see also Fig.~S6 in supplements).

To numerically study the few-photon regime, we start with the uncoupled case ($g=0$) and prepare the matter in its ground state and the cavity in a photon-number state ($\ket{n=N_{ph}}$). The coupling $g$ is then turned on \emph{adiabatically}. In the non-resonant regime ($U\ne \Omega$), the system approaches the lowest-lying state within the subspace $n=N_{ph}$ due to the adiabatic theorem (generally an \emph{in-state} of the quantum scattering problem). We solve the time-evolution of the cavity-coupled Hubbard chain of $L=8$ using a Krylov-space algorithm for iteration number $80$. The coupling $g$ is raised to a very large value $g=1.0$ for demonstration.

\begin{figure}
\includegraphics[scale=1.35]{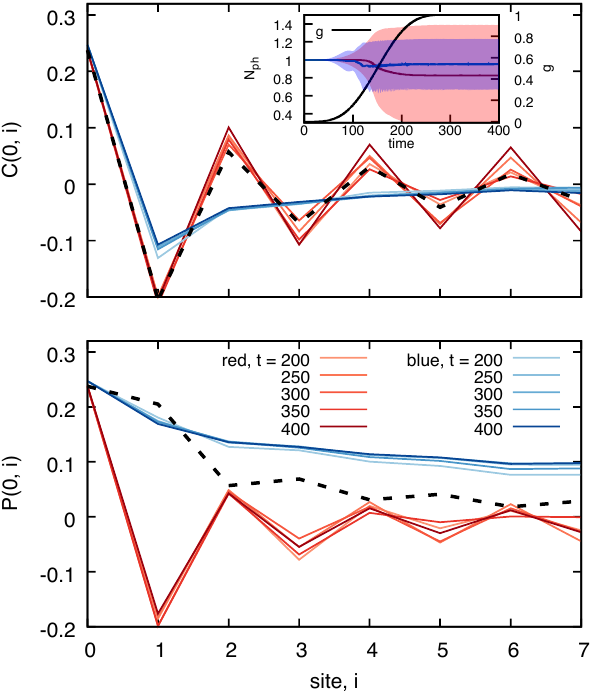}
\caption{The evolution of charge and pairing correlations under the injection of one photon. The red curves represent the 
red-detuned case $\Omega = 0.5U$ while the blue curves represent the blue-detuned case $\Omega = 1.5U$. The 
dashed line represents the initial state $g=0.0$ (ground state without light-matter coupling). The inset shows the quench profile of coupling $g$ from $0.0$ to $1.0$ and the evolution of photon number in the cavity. The shaded area covers the region of $N_{ph}(t)\pm\Delta N_{ph}(t)$ where $\Delta N_{ph}=\langle a^\dag a a^\dag a \rangle-\langle a^\dag a \rangle^2$ is the uncertainty of photon number.}
\label{quench}
\end{figure}

The result is shown in Fig.~\ref{quench}. As $g$ is turned on, the photon number drops from $N_{ph}=1$ to $0.953$ for $\Omega=1.5U$ (blue-detuned) and to $0.826$ for $\Omega=0.5U$ (red-detuned). In contrast to the ground state, the CDW becomes significantly suppressed while the SC is enhanced for the blue-detuned cavity. Thus, an entangled photon-order state distinct from the equilibrium is dynamically created by driving. In the red-detuned case the CDW order is again enhanced, but, instead of a strong suppression, the pairing correlation is turned into a staggered form, i.e., the $\eta$--pairing SC \cite{kaneko2019,kaneko2019b,li2020prb}. 

\textbf{Remarks on the BCS limit. }So far we have concentrated on the strong coupling or BEC limit \cite{micnas1990}, where projecting out higher excited states is justified. One can also study the weak interacting or BCS ($U\sim t_0$) regime in the limit $\Omega\gg t_0$, which results in the simplified $H^{\rm eff}_{nn}=\mathcal{H}_{nn}+\sum_l \mathcal{H}^0_{n,n+l}\mathcal{H}^0_{n+l,n}/l\Omega$. In particular, this induces a next-to-nearest-neighbor (NNN) hopping and a two-site interaction 
$I_1(\eta^+_i\eta^-_j+{\rm h.c.})+2I_2\eta^z_i\eta^z_j+2I_2\bm S_i\cdot \bm S_j$, with coefficients $I_1=\sum_{l>-n,l\ne0}j_{n,n+l}j_{n+l,n}/l\Omega$ and $I_2=\sum_{l>-n,l\ne0}(-1)^l j_{n,n+l}j_{n+l,n}/l\Omega$, leading to qualitatively similar physics as described in the BEC limit. Note that the cavity also induces a long-range interaction close to the ground state, which is, for the lowest order, of current-current type \cite{schlawin2019}. This may complicate the scenario in certain parameter regimes. A systematic examination in this regime as well as the BCS--BEC crossover is reserved for the future.

Finally, we comment on the experimental realization. Our findings are related to recent experiments on correlated materials \cite{keimer2015,battisti2017} coupled to surface plasmon polaritons, and the cavity-control of SC order \cite{thomas2019}. An interesting material class is organic charge transfer salts, in which superconductivity competes with various orders \cite{Dumm2009,buzzi2020}. With a charge transfer energy $U\sim 1eV$, assuming $\Omega\sim U$ gives a cavity wavelength $\lambda_c\sim 1\mu\mathrm{m}$. To reach $g \gtrsim 0.1$, one needs an effective cavity volume $V/\lambda_c^3\lesssim 10^{-7}$. Recent experimental advances hold the promise to reach this parameter regime \cite{kockum2019}. We further propose two routes to test our conclusions: (i) Artificial systems with dynamical $U(1)$ gauge fields, e.g., proposed in recent cold-atom experiments, can be minimally described by Hamiltonian~\eqref{Ho} \cite{mazurenko2017,mil2020}. (ii) Solid-state systems coupled to a \emph{continuum} of photon modes can be realized with a Fabry-Perot cavity \cite{rokaj2020}. The contributions from all modes add up cooperatively, yielding an effective coupling not limited by the cavity volume \footnote{The Fabry-Perot cavity can be implemented, for example, by fabricating a Van der Waals heterostructure consisting of metallic mirrors and the layered material. A detailed analysis has been carried out and will be published elsewhere.}. This setting should still allow for a selective enhancement of CDW and SC, controlled by ratio of $U$ and the cavity frequency.

\textbf{Conclusion.}  
In this paper, we demonstrate the concept of controlling competing orders using quantum light with a minimal model of competing CDW and SC orders, the cavity-coupled attractive Hubbard model, solved by an analytic theory based on the quantum Floquet formalism, and then confirmed by exact diagonalization for 1D chains. The vacuum fluctuations become entangled with the electronic ordering and enhance exclusively the CDW order, giving rise to a putative superradiant condenstate for large system sizes. This differs dramatically from the Floquet-engineering scenarios. By injecting few photons in the cavity, one can furthermore selectively enhance different orders, including CDW, $s$--wave SC, and $\eta$--pairing SC in different parameter regimes. 

The quantum Floquet formalism provides a natural framework to unify the quantum driving and the classical Floquet scenarios \cite{kitagawa2011}, and can be combined with established numerical methods, such as dynamical mean-field theory and its extensions \cite{tsuji2008,aoki2014,golez2019} to describe more complicated systems \cite{curtis2019,claassen2017}, such as driven and open cavities \cite{lentrodt2020, gao2020}. 

\begin{acknowledgments}
We thank M.~A.~Sentef for useful discussions. This work was supported by ERC Starting Grant No. 716648.
\end{acknowledgments}

\bibliography{corder.bib}
\appendix
\renewcommand{\theequation}{S.\arabic{equation}}
\renewcommand{\thefigure}{S.\arabic{figure}}

\widetext
\clearpage
\begin{center}
\textbf{Supplemental material for ``Manipulating intertwined orders in solids with quantum light"}
\end{center}

\section{The evaluation of quantum Floquet matrix Hamiltonian}
In this section we show the evaluation of $\bra{n}e^{i\hat{\phi}_{ij}}\ket{m}$. We use the Baker-Hausdorff formula $\exp(X+Y)=\exp(X)\exp(Y)\exp(-[X,Y]/2)$ when $[X,Y]$ is a c-number, then Taylor expand the exponential factor and reorder the sum, 
\begin{align}
	e^{i\xi_{ij}g(a + a^\dag)}&=e^{i\xi_{ij}ga^\dag} e^{i\xi_{ij}g a} e^{-g^2/2}\nonumber\\
&=e^{-g^2/2}\sum_{kk'}\frac{(ig\xi_{ij}a^\dag)^k(ig\xi_{ij}a)^{k'}}{k!k'!}\nonumber\\
&=e^{-g^2/2}\left[\sum_{k}\frac{(-1)^k g^{2k}}{k!k!}(a^\dag)^{k} a^{k} + \sum_{l>0}\sum_{k}\frac{(-1)^k g^{2k}}{k!(k+l)!}\left((ig\xi_{ij})^l(a^\dag)^{k+l} a^{k} + (ig\xi_{ij})^l(a^\dag)^k a^{k+l}\right)\right]\nonumber\\
	&=\sum_l i^{|l|}\xi_{ij}^{l}\oJ_l.
\end{align}
Here we have defined 
\begin{align}
	\oJ_l\equiv e^{-g^2/2}\sum_k\frac{(-1)^k g^{2k + |l|}}{k!(k+|l|)!}
	\begin{cases} (a^\dag)^{k+|l|}a^k, &l\ge0\\ 
	(a^\dag)^k a^{k+|l|}, &l<0\end{cases}.
	\label{oJ}
\end{align}
This functional satisfies $\oJ_l^\dag=\oJ_{-l}$. We can, therefore, evaluate the quantum Floquet Hamiltonian and define $j_{n,m}=\bra{n}\oJ_{n-m}\ket{m}$. For $n>m$ one obtains
\begin{align}
	j_{n,m} = e^{-g^2/2}\sum_{k=0}^m\frac{(-1)^kg^{2k+|n-m|}}{k!(k+|n-m|)!}\sqrt{\frac{n!}{m!}}\frac{m!}{(m-k)!},
\end{align}
and similarly for $n<m$
\begin{align}
	j_{n,m} = e^{-g^2/2}\sum_{k=0}^n\frac{(-1)^kg^{2k+|n-m|}}{k!(k+|n-m|)!}\sqrt{\frac{m!}{n!}}\frac{n!}{(n-k)!}.
\end{align}
The $j_{n,m}$ function is a finite sum which can readily be evaluated. As discussed in the main text, for $|n-m|\to\infty$, $j_{n,m}$ decays 
as $g^{|n-m|}/\sqrt{|n-m|!}$ so that the coupling between different quantum Floquet bands quickly decays to zero as $|n-m|$ increases.

\section{The cavity to Floquet crossover in the semiclassical limit}
\begin{figure}
\includegraphics[scale=1.3]{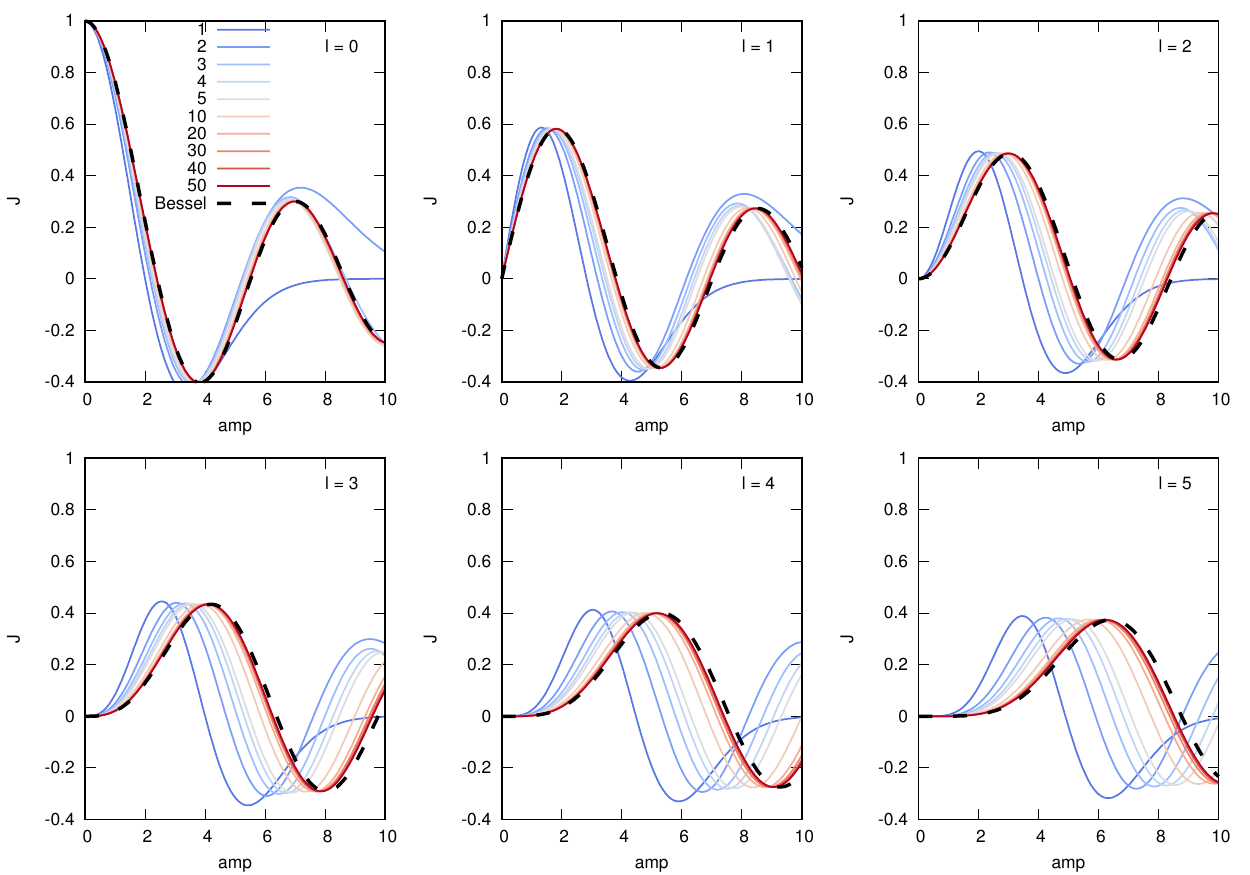}
\caption{The convergence of $j_{n,m}$ to the Bessel function $J_l(\mathcal{A})$ under the semiclassical limit. The amplitude is $\mathcal{A}=2g\sqrt{n}$. }
\label{jnm}
\end{figure}

\begin{figure}
	\includegraphics[scale=0.7]{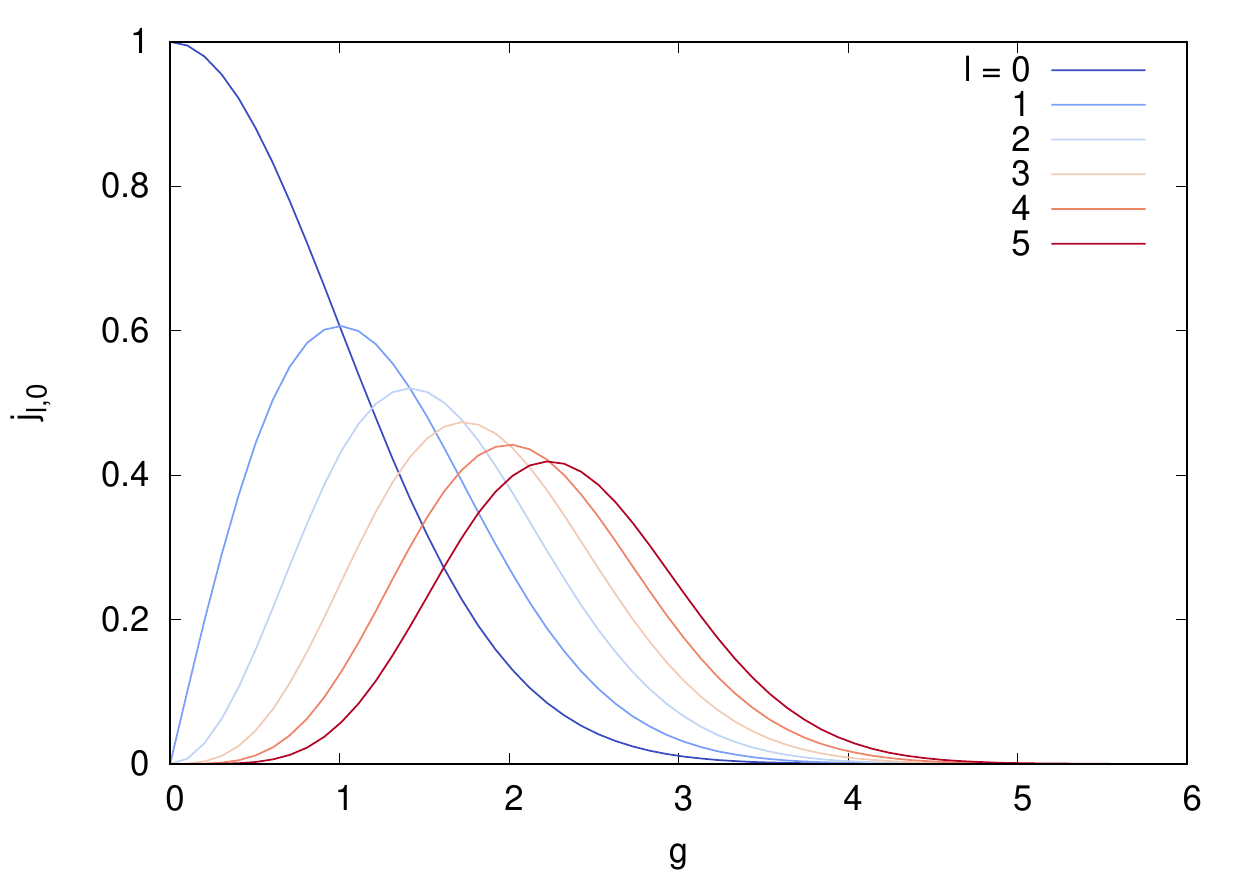}
	\caption{The function $j_{l,0}(2|g|)$ for the dark cavity. }
	\label{jl0}
\end{figure}

In this section we explicitly show how the quantum Floquet Hamiltonian continuously converges to the Floquet Hamiltonian. This is essentially the crossover from quantum driving by fluctuations to semiclassical driving by a coherent classical field. We will follow the physical intuition of Ref.~\citenum{sentef2020}. Under periodic driving, where the \emph{classical} vector potential $\mathcal{A}\cos \Omega t$ coupled through the Peierls phase $e^{i\xi_{ij}\mathcal{A}\cos \Omega t}$, the corresponding Floquet matrix Hamiltonian reads \cite{tsuji2008},

\begin{align}
 \mathcal{H}^F_{nm}&=-t_0\sum_{\langle ij\rangle\sigma} i^{|n-m|}\s_{ij}^{n-m}J_{|n-m|}(\mathcal{A})c^\dag_{i\sigma} c_{j\sigma} + \left(-U \sum_{i} 
n_{i\uparrow}n_{i\downarrow}
+n\Omega\right)\delta_{nm},
\end{align}

where $J_l(x)$ is the $l$th Bessel function of the first kind,
\begin{align}
J_{|l|}(x) = \sum_k (-)^k\frac{(x/2)^{2k+|l|}}{k!(k+|l|)!}.
\end{align}
Back to our quantum Floquet formulation, in the semiclassical limit $n\to\infty$ and $g\to0$, with $2g\sqrt{n}=\mathcal{A}$. Suppose $l\ge0$, the $j_{n+l,n}$ function reads
\begin{align}
	\lim_{n\to\infty}j_{n+l,n}=\lim_{n\to\infty}e^{-\mathcal{A}^2/8n}\sum_{k=0}^n\frac{(-1)^k(|\mathcal{A}|/2)^{2k+|l|}}{k!(k+|l|)!}\sqrt{\frac{(n+|l|)!}{n!n^{|l|}}}\frac{n!}{(n-k)!n^k}
	=J_{|l|}(\mathcal{A}).
\end{align}
Note that $(n+\ell)!/n!\to n^\ell$ under the limit. This restores the Floquet-driven case where the vector potential $\bm{A}\cdot\bm{d}_{ij}$ is replaced by a Peierls phase $\xi_{ij}\mathcal{A}\cos(\Omega t)$. This limiting behavior is shown in Fig.~\ref{jnm}. 
In the ground state the relevant photon-number sector is $n=0$, the $j_{l,0}$ functions reduce to $j_{l,0}=e^{-g^2/2}g^l/\sqrt{l!}$, which are plotted in Fig.~\ref{jl0}. For fixed $g$, $j_{l,0}$ simply gives the weight of the photon state $\ket{l}$ after the Peierls phase acting on the vacuum state, which leads to a coherent state, where 
$|j_{l,0}|^2$ is a Poisson distribution.

\section{The light-matter coupling in the effective model}
Under the strong-coupling expansion, the unperturbed electronic states $\ket{s}$'s (eigenstates of $H_U$) mix with different photon-number sectors according to the second-order perturbation theory in $t_0/U$. This introduces entanglement between electrons and the quantum light. Furthermore, when the CDW order is considered, one obtains further effective coupling between CDW and the electric field $\bm E= ig\Omega \bm e_p(a-a^\dag)$.

As a heuristic method, we compute approximately the effective Hamiltonian at the strong $U$ limit ($U\gg \Omega, t_0$). By projecting out electronic excitations, the effective Hamiltonian within the $n$th photon-sector reads
\begin{align}
H^{\rm eff}_{n+1,n}&=\sum_{m} \mathcal{P}_0\mathcal{H}^0_{n+1, m}\mathcal{P}_1\mathcal{H}^0_{m,n}\mathcal{P}_0/U\nonumber\\
&\approx \mathcal{P}_0\mathcal{H}^0_{n+1, n}\mathcal{P}_1\mathcal{H}^0_{n,n}\mathcal{P}_0/U+\mathcal{P}_0\mathcal{H}^0_{n+1, n+1}\mathcal{P}_1\mathcal{H}^0_{n+1,n}\mathcal{P}_0/U\nonumber\\
&\approx g\frac{2t_0^2}{U}i\sqrt{n+1}\sum_{\langle ij\rangle}\xi_{ij}(\eta^z_i-\eta^z_j)
\end{align}
and similarly for $H^{\rm eff}_{n,n+1}$. To restore the photon operators, we re-sum $H^{\rm eff} \approx \sum_n (H^{\rm eff}_{n+1,n}\otimes\ket{n+1}\bra{n}+{\rm h.c.})$ and identify $a^\dag=\sum_n\sqrt{n+1}\ket{n+1}\bra{n}$. Using the notation of (4) in the main text, one then obtains a light-matter coupling term with the form $ig_{\rm eff}(a-a^\dag)\sum_{\langle ij\rangle}\xi_{ij}(\eta^z_i-\eta^z_j)$. Recall $\eta^z_i=(n_i-1)/2$, the term turns out to be $ig_{\rm eff}(a-a^\dag)(n_0-n_{L-1})$ for an open Hubbard chain of site number $L$. This term should be responsible for the double-peak structure shown in Fig.~2 of the main text.

There is another linear coupling term, which couples {$\bm A$} with pair current {$i(\eta^x_i\eta^y_j-\eta^y_i\eta^x_j)$}, but $ig_{\rm eff}(a-a^\dag)\sum_{\langle ij\rangle}\xi_{ij}(n_i-n_j)$ should be the relevant term in the regime where CDW fluctuations are enhanced by the coupling to the cavity.

\section{Nonlinearity in the photon states}
\begin{figure}
	\includegraphics[scale=0.4]{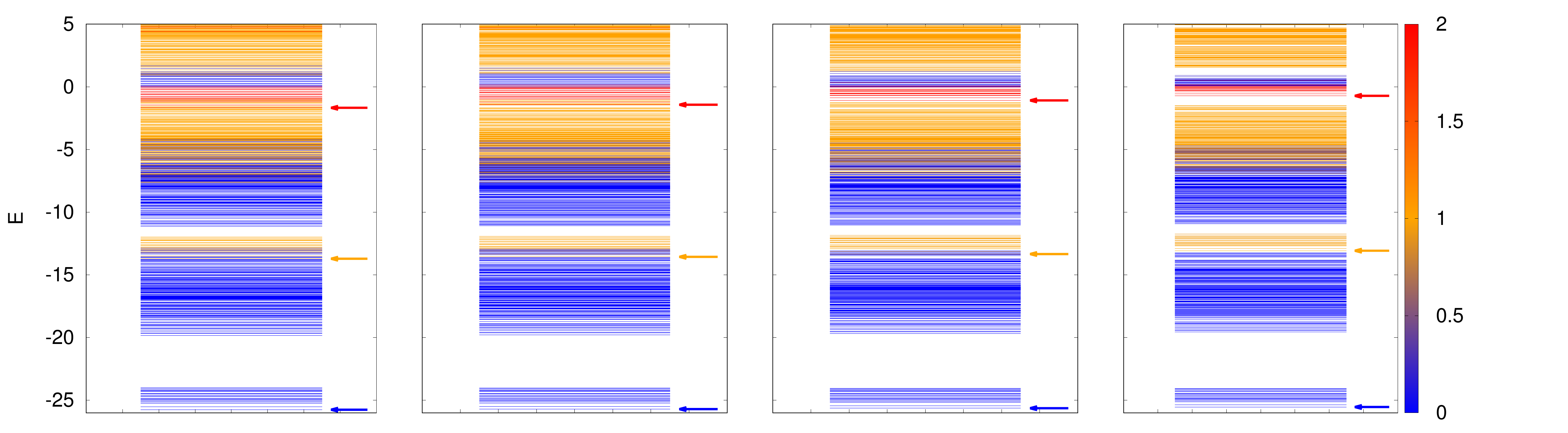}
	\caption{The energy spectrum of $L=6$ Hubbard chain coupled to the blue-detuned cavity ($\Omega=1.5U$). The color represents the photon number $N_{ph}$. The arrows label the eigenstates picked up in the Fig.~\ref{phlevel}. The four panels correspond to $g=0.1,0.2,0.3,0.4$ from left to right, respectively.}
	\label{e_blue}
\end{figure}

\begin{figure}
	\includegraphics[scale=0.4]{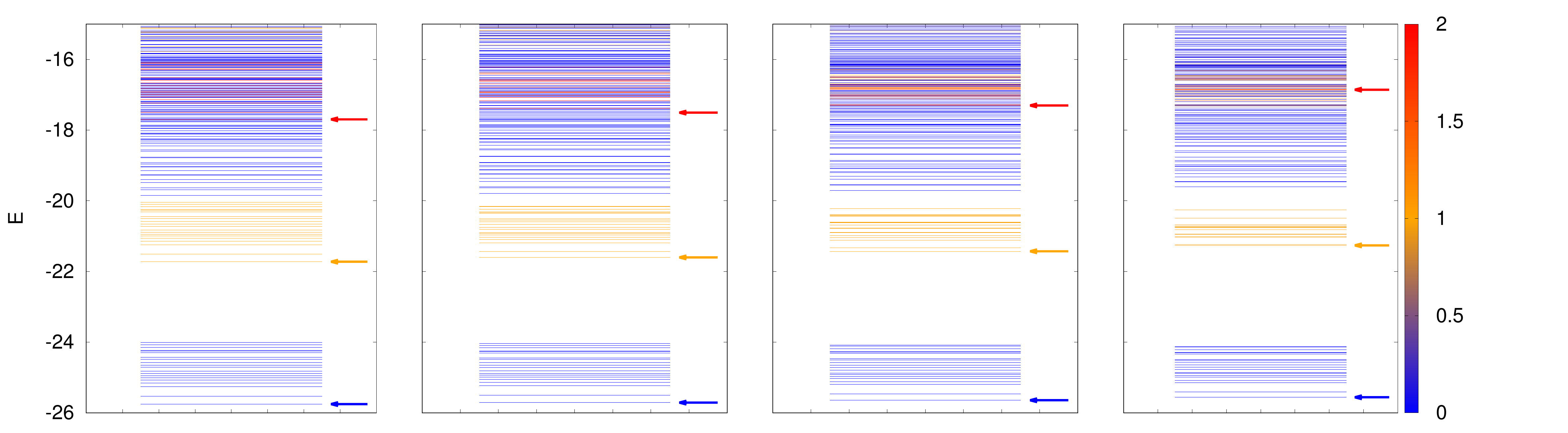}
	\caption{The energy spectrum of $L=6$ Hubbard chain coupled to the red-detuned cavity ($\Omega=0.5U$). The color represents the photon number $N_{ph}$. The arrows label the eigenstates picked up in the Fig.~\ref{phlevel}. The four panels correspond to $g=0.1,0.2,0.3,0.4$ from left to right, respectively.}
	\label{e_red}
\end{figure}

To demonstrate the nonlinearity in the photon spectrum, we have computed all of the eigenvalues for the cavity-coupled Hubbard chain of $L=6$. In this section, we give more details on the energy spectrum. The energy eigenvalues are shown in Fig.~\ref{e_blue} for the blue-detuned cavity and in Fig.~\ref{e_red} for the red-detuned cavity. In the blue-detuned case, due to the relatively larger $\Omega$, the few-photon states are rather protected by an energy gap and the photon number $N_{ph}=\langle a^\dag a\rangle$ of the excited states is relatively discrete, taking values around the integers. Interestingly, the first photon gap $0\to1$ seems less robust, due a photoemission-like process where mobile electronic excitations are formed upon the absorption of a photon. 

\begin{figure}
\includegraphics[scale=0.7]{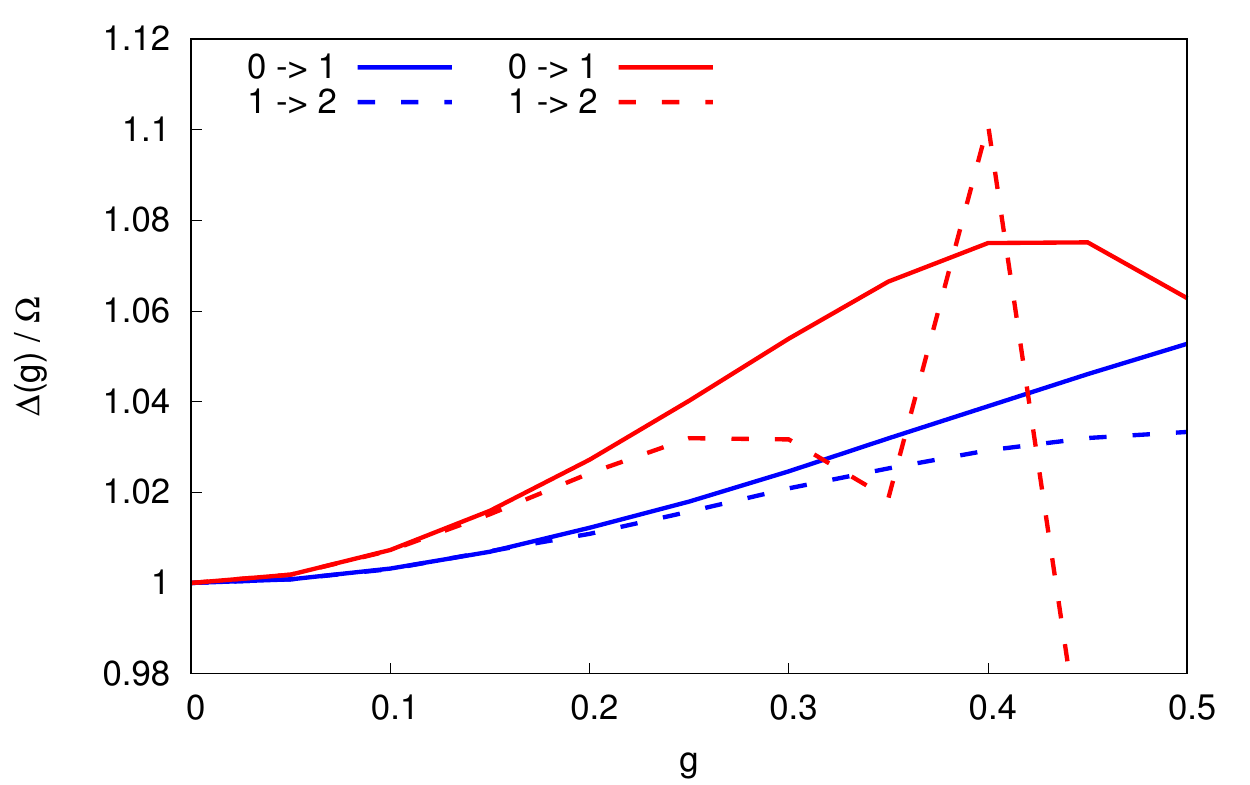}
\caption{The shift in photon energy gaps due to light-matter coupling. The energy gaps are obtained by picking up the lowest-lying eigenstates with photon-number $n=0,1,2$ and evaluating the energy difference for an $L=6$ cavity-Hubbard chain. The red (blue) curves correspond to $\Omega/U=0.5$ ($\Omega/U=1.5$). In the red-detuned regime, the $n=2$ sector strongly mixes with electronic excitations, resulting in dramatic changes in the $1\to 2$ curve for larger coupling.}
\label{phlevel}
\end{figure}

On the other hand, for the red-detuned case, there is a clear gap between $N_{ph}=1$ and $N_{ph}=0$ states, since the energy of one photon is well below $U$. However, the two-photon state appears to strongly mix with the electronic excitations, leading to superpositions of different photon-number sectors. Strictly speaking, in the large coupling regime $g\sim0.4$, the $1\to2$ photon gap is not completely well-defined because $U=2\Omega$ indeed satisfies the resonance condition. This results in the dramatic change in the $1\to2$ curve in Fig.~\ref{phlevel}. After all, the one-photon engineering regime appears to be well-defined, and should be accessible from an adiabatic injection of photons into the cavity.

By picking up the lowest-lying states in each photon-number sector, we show the effective photon gap in Fig.~\ref{phlevel}. In practice, due to the mixing between different photon-number states, we have actually set a photon number threshold $N_{ph}=0.5$ ($1.5$) for the one-photon (two-photons) state. This turns out to give reasonable results at least for not-too-large coupling $g$.

\section{Enhanced superconductivity under external driving}
Despite the photon nonlinearity shown in the last section, creating a real \emph{Fock state} in experiments can still be challenging. In this section, we demonstrate the enhanced SC correlation is indeed robust for the few-photon regime even for non-Fock states by considering a continuous external driving. 

\begin{figure}
\includegraphics[scale=1.5]{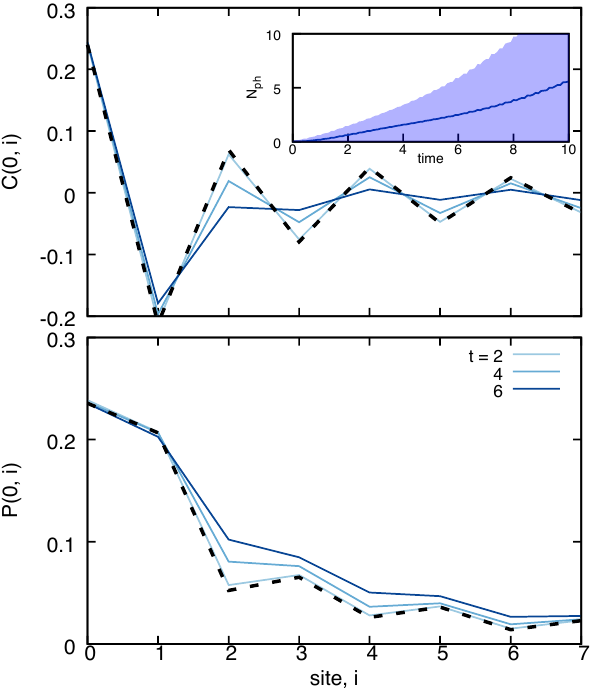}
\caption{The enhancement of SC under continuous driving. The inset of panel (a) shows the time evolution of average photon number $N_{\rm ph}=\langle a^\dag a\rangle$ and its standard deviation (blue shades). }
\label{driving}
\end{figure}

For simplicity, we consider the lossless single-mode cavity described in the main text. In particular, a blue detuned cavity $\Omega=1.5U$ with $U=8.0$ and $g=0.5$ is considered to study the enhancement of SC. An external driving term $F_0\sin(\Omega_{\rm dr}t)$ is applied to the cavity-matter system. We choose $F_0=1.0,\Omega_{\rm dr}=1.5U$ to resonantly excite the cavity mode. The results are shown in Fig.~\ref{driving}. The pairing correlation is clearly enhanced as the photon number $N_{\rm ph}$ continuously increases. The CDW correlation is also suppressed as predicted by the theory. It is worth noting that the photon number exhibits a large variance (indicated by the blue shades), showing the photon state deviates from a Fock state. However the enhancement of SC is still robust. 

The dynamics shown in Fig.~\ref{driving} is within tens of hopping times, or roughly speaking femtoseconds. The dissipation from an open cavity is often negligible in this time regime, and the closed-cavity approximation is justified. For longer times the dissipation of cavity should be taken into account, which may better stabilize a photon Fock state in the nonequlibrium steady-state.

\end{document}